# Data Format Standardization and DICOM Integration for Hyperpolarized $^{13}$C MRI

Ernesto Diaz[1], Renuka Sriram[1], Jeremy W. Gordon[1], Avantika Sinha[1], Xiaoxi Liu[1], Sule Sahin[1], Jason Crane[1], Marram P Olson[1], Hsin-Yu Chen[1], Jenna Bernard[1], Daniel B. Vigneron[1,2], Zhen Jane Wang[1], Duan Xu[1,2], Peder E. Z. Larson[1,2]

Affiliations:

[1]Department of Radiology and Biomedical Imaging, University of California – San Francisco, San Francisco, California, USA

[2]UC Berkeley-UCSF Graduate Program in Bioengineering, University of California, Berkeley and University of California, San Francisco, California, USA

Corresponding author address:

peder.larson@ucsf.edu, 1700 4th Street, Byers Hall Room 102C, San Francisco, CA  94143

# Statements and Declarations

**Funding**

This work was supported by NIH grants P41EB013598, U24CA253377, R01CA262630, and R01CA249909. Authors PEZL, DBV, and JWG received research support from GE Healthcare.

**Competing Interests**

Financial interests: Authors ED, RS, AS, XL, JB, ZJW, and XU declare they have no financial interests. Authors PEZ, DBV, and JWG received research funding from GE Healthcare

**Author Contributions**

All authors contributed to the study conception and design. DICOM tools were created by Ernesto Diaz, Jeremy Gordon, Jason Crane, Marram P Olson, and Xiaoxi Liu.  Data collection and analysis were performed by Ernesto Diaz. Creation of the proposed parameters was performed by Peder Larson.  The first draft of the manuscript was written by Ernesto Diaz and Peder Larson, and all authors commented on previous versions of the manuscript. All authors read and approved the final manuscript.

**Ethics approval**

This manuscript presents technical methods but does not include any in vivo study results.

**Consent to participate**

Not applicable

**Consent to publish**

Not applicable

# Abstract

Hyperpolarized (HP) $^{13}$C MRI has shown promise as a valuable modality for *in vivo* measurements of metabolism and is currently in human trials at 15 research sites worldwide. With this growth it is important to adopt standardized data storage practices as it will allow sites to meaningfully compare data.

In this paper we (1) describe data that we believe should be stored and (2) demonstrate pipelines and methods that utilize the Digital Imaging and Communications in Medicine (DICOM) standard. This includes proposing a set of minimum set of information that is specific to HP $^{13}$C MRI studies. We then show where the majority of these can be fit into existing DICOM Attributes, primarily via the "Contrast/Bolus" module.

We also demonstrate pipelines for utilizing DICOM for HP $^{13}$C MRI. DICOM is the most common standard for clinical medical image storage and provides the flexibility to accommodate the unique aspects of HP $^{13}$C MRI, including the HP agent information but also spectroscopic and metabolite dimensions. The pipelines shown include creating DICOM objects for studies on human and animal imaging systems with various pulse sequences. We also show a python-based method to efficiently modify DICOM objects to incorporate the unique HP $^{13}$C MRI information that is not captured by existing pipelines. Moreover, we propose best practices for HP $^{13}$C MRI data storage that will support future multi-site trials, research studies and technical developments of this imaging technique.

# Introduction

Hyperpolarized (HP) $^{13}$C MRI has shown promise as a valuable modality particularly for *in vivo* measurements of metabolism[1], and is currently in human trials at 15 research sites worldwide with over 75 human research publications[2]. It is based on intravenous injection of a hyperpolarized contrast agent that has been enriched with $^{13}$C. The most widely used agent is $^{13}$C-pyruvate for measuring metabolism, but there are also a range of other promising agents undergoing clinical translation including $^{13}$C-urea (perfusion)[3], $^{13}$C-alpha-ketoglutarate (metabolism)[4], and $^{13}$C-fumarate (necrosis)[5]. To date, human studies have been performed to characterize normal metabolism, and clinical studies have been performed to study in altered metabolism in prostate cancer, brain tumors, kidney tumors, pancreatic cancer, metastatic disease, liver disease, and heart disease[2].

As HP $^{13}$C MRI continues to grow, it is important to adopt standardized data storage practices, allowing sites to compare data. The Digital Imaging and Communications in Medicine (DICOM) format is an attractive standard format as it is commonly used in digital imaging, medical imaging, and communications. The DICOM file type consists of the image and the metadata of the image packed into a single file. The information in the metadata is organized as a constant and is standardized by a series of data elements. By extracting the data elements, we can access important information regarding the patient demographics and study parameters that are crucial for study interpretation. The DICOM standard is actively supported and updated. It also supports the interchange of information between computer systems such as picture archiving and communication system (PACS).

The goal of this work is to describe an approach for HP $^{13}$C MRI data standardization utilizing the DICOM format. We first describe the data requirements, focusing on the unique aspects of HP $^{13}$C MRI. We then describe current processing pipelines that support the creation of DICOM objects from multiple vendors and pulse sequences. Finally, we demonstrate how this information can be incorporated into DICOM format.

# Hyperpolarized $^{13}$C MRI Data

The data requirements for HP $^{13}$C MRI are outlined briefly below. This includes performing imaging to spatially resolve signals, encoding of multiple metabolites to measure metabolic conversion, performing dynamic measurements to capture the rapid kinetics, and characteristics of the HP agent, all of which that will affect the results and analysis.

## Imaging

HP $^{13}$C experiments can be spatially resolved using MRI techniques, providing important localization to visualize metabolic activity within the body. Therefore, a format to store multi-dimensional data needs to be supported.

## Metabolite encoding

HP $^{13}$C MRI has the unique ability to encode signals from multiple metabolites, distinguishing the injected agent, or substrate, from resulting metabolic products that are converted in vivo. The most common combination of metabolites is [1-$^{13}$C]pyruvate (substrate), [1-$^{13}$C]lactate (product), [1-$^{13}$C]alanine (product), and $^{13}$C-bicarbonate (product). Thus, the data should support a metabolite encoding dimension. Metabolite encoding in the acquisition is typically done with MR spectroscopy methods, chemical shift encoding (e.g. IDEAL) methods, or metabolite-specific imaging[6]. The data must support metabolite encoding through identification of the individual compounds and/or the resonant frequencies, or with a spectroscopy dimension.

### Dynamic Measurements

In a HP $^{13}$C MRI experiment, rapid metabolite kinetics following injection, including perfusion and metabolic conversion, that are typically captured with time-resolved imaging. Thus, the data should support dynamic measurements and record timings of the measurements.

### HP $^{13}$C Agents

To analyze HP $^{13}$C data, it is important to know the characteristics of the HP $^{13}$C agent, how it was prepared, and how it was used. Table 1 lists such relevant HP $^{13}$C agent information. Thus, the data should support capturing of this information.

| HP $^{13}$C Agent Information | Purpose/Description |
| --- | --- |
| **HP $^{13}$C agent(s) injected** | Describe the composition of the injected agents. Note that this can include simultaneous injection of multiple agents |
| **Administration route of agent** | How was the HP solution delivered |
| **Total Injected Volume** | How much HP solution was injected in total |
| **Concentration of each HP agent** | Concentration of each HP agent in HP solution |
| **Start Time of injection** | Describe the timing of the start of injection, important to know relative to the data acquisition timing |
| **Total injection duration** | Over how long was the HP solution injected |
| **Polarization** | What was the measured polarization of the HP agent(s). Typically back-calculated to the time of dissolution. |
| **Polarization Measurement Timing** | timing of the polarization measurement relative to the data acquisition |
| **Agent Relaxation Rates (e.g. T1)** | Expected or measured relaxation rates used to calculate polarization |
| **Dissolution Timing** | Timing of the dissolution of the HP agent, most important to know relative to the data acquisition |

Table 1: Hyperpolarized $^{13}$C agent information that we typically keep track of during our studies.

## DICOM Processing Pipelines

Our current processing pipelines support DICOM creation from multiple vendors and pulse sequences, listed in Table 2. The spectroscopy sequences use an open-source standards-based software framework(SIVIC) for DICOM creation which includes support for DICOM Spectroscopy (7,8).

| Scanner Vendor | Sequences | Raw Data | Tools | Output |
|---|---|---|---|---|
| Bruker | Slab-selective MRS, chemical shift imaging (CSI) and Echo-planar spectroscopic imaging | MRS Raw File or Dad File | SIVIC  MATLAB | MR Spectroscopy DICOMs |
| Bruker | Metabolite-Specific Imaging | FID | Bruker Paravision  MATLAB | Metabolite image, AUC, kinetic rate DICOMs |
| GE Healthcare | Slab-selective MRS, chemical shift imaging (CSI) and Echo-planar spectroscopic imaging | Pfile or ScanArchive | SIVIC | MR Spectroscopy DICOMs |
| GE Healthcare | Metabolite-specific EPI | Pfile or ScanArchive | GE Orchestra Toolbox  MATLAB | Metabolite images, AUC ratio and kinetic rate DICOMs |
| RTHawk Research (Vista.ai) | Metabolite-specific spiral and bSSFP | RTHawk raw data  Real-time reconstructed DICOM | MATLAB to improve data | Metabolite image DICOMs |

Table 2: Currently supported sequences with DICOM creation pipelines in use at our institution in alphabetical order.

## MR Spectroscopy and Spectroscopic Imaging with SIVIC

MR spectroscopy and spectroscopic imaging (MRS/I) for HP $^{13}$C MRI includes data sampling at various time delays to resolve a spectrum and separate $^{13}$C-labeled compounds. For processing of MRS/I data, we utilize the libraries in SIVIC for data processing and DICOM creation, including support of the DICOM MRS standard[9]. This software can reads in multiple vendor MRS raw data files (Bruker, GE, Philips, Siemens, Varian), and it for HP $^{13}$C MRI studies on GE Healthcare and Bruker MRI scanners.

1. Reads in raw data
2. Performs MR spectroscopy or MR spectroscopic imaging reconstructions

3. Writes to DICOM MRS file
4. Optional – extract peak amplitudes of areas from spectra and create individual metabolite DICOM files
5. Optional - write raw data or partially processed data to DICOM
6. Optional - secondary capture of MRS and anatomical visualization to DICOM

## GE Metabolite-specific Imaging

Metabolite-specific imaging for HP $^{13}$C MRI uses a spectral-spatial RF pulse to selectively excite a single metabolite[10] and is followed by fast imaging readout such as echo-planar imaging (EPI) or spirals. We have implemented EPI-based metabolite-specific imaging for GE Healthcare MRI scanners along with a reconstruction pipeline based on the GE Orchestra reconstruction toolbox[11]. This leverages the EPI processing and DICOM creation functionality that is utilized by GE product pulse sequences.

1. Reads in raw data (Pfile or ScanArchive)
2. Perform EPI reconstruction including Nyquist ghost correction
3. Perform HP-specific coil combination[12]
4. Create area-under-curve (AUC) and kinetic rate maps
5. Write metabolite images and parameter maps to DICOM

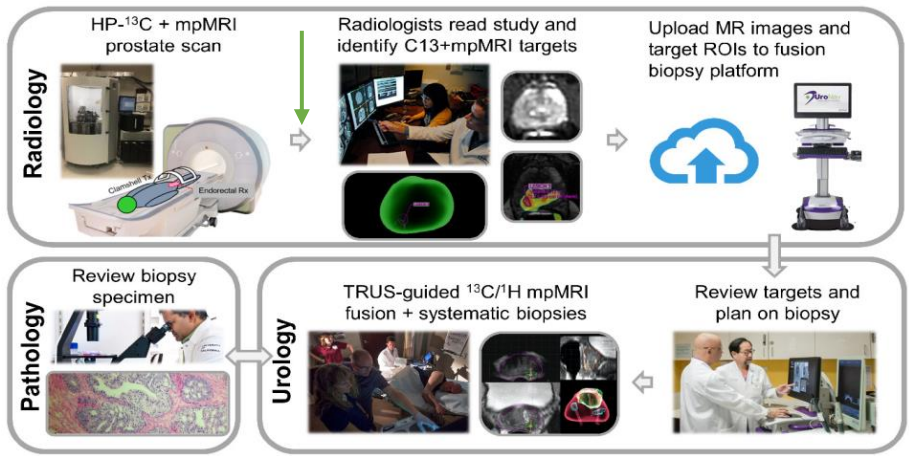

Figure 1: Example of how HP $^{13}$C MRI can be used in a workflow for improved targeting of biopsies in prostate cancer[11]. Using DICOM has been critical in the success of this workflow, as this allows for integration into PACS as well as the fusion biopsy platform, with radiologists and urologists easily able to review the images and resulting annotations outlining the biopsy targets. Figure reproduced with permission from[11].

## RTHawk Metabolite-specific Imaging

We have implemented metabolite-specific imaging with EPI and spirals, as well as metabolite-specific balanced steady-state free-precession (bSSFP) sequences[13-15] on the RTHawk Research platform. This is a vendor-neutral platform that allows for advanced control of the MRI system including real-time reconstructions and feedback as well as pulse sequence programming.

1. We utilize real-time reconstruction built into RTHawk during the scanning process, resulting in individual metabolite DICOM images at each timepoint.(Note: This uses the sum of squares coil combination technique.)
2. The DICOMs Images are exported when the data is transferred from the RTHawk workstation to the local network.
3. A second, improved image reconstruction is performed using raw complex-valued data and a more advanced coil combination method to enhance the data quality[9].
4. To integrate the improved data into DICOM, a MATLAB script is employed that reads the metadata from the real-time reconstructed DICOMs Images, replaces the original voxel data with the results of the second, optimized reconstruction, and creates new DICOMs Images with appropriate new unique identifiers (UIDs).

### Bruker Metabolite-specific Imaging

Similar to the GE metabolite-specific imaging sequence, it uses a spectral-spatial RF pulse to excite a single metabolite[10], and is followed by a fast echo-planar imaging (EPI) readout[16].

1. Bruker scanner has its own built-in EPI reconstruction and an option to write reconstructed metabolite maps to DICOM files.
2. A MATLAB script is used to read in the DICOM files. From that data, we generate AUC maps and kinetic rate maps, and write back to DICOM files.
3. Optional: Reads in Bruker raw data using a MATLAB script, use a custom reconstruction script and write to DICOM.

## DICOM HP Agent Integration

Many of the HP agent metadata that are desirable for analyses (Table 1) can be integrated into existing DICOM attributes, particularly within the "Contrast/Bolus" module. However, the above DICOM pipelines by default do not include support for these metadata. To address this, we built a tool to modify HP $^{13}$C DICOM objects based on user input to include this HP agent metadata.

Currently, the HP $^{13}$C images created by MRI scanners or in-house reconstruction pipelines do not utilize the Contrast/Bolus Attributes. We seek to leverage the existing Contrast/Bolus Attributes in DICOM to store information to identify the study and experimental parameters that are valuable for HP $^{13}$C data analysis by researchers and clinicians. Table 3 lists the attributes we are adding to our images.

| TAG | VR | Attribute Name | $^1$H MRI Example Values | HP $^{13}$C Description | HP $^{13}$C Example Values |
|---|---|---|---|---|---|
| (0018,0010) | LO | Contrast/Bolus Agent | CC MAGNEVIST<br><br>ORAL & OMNIPAQUE | HP $^{13}$C agent(s) injected | HYPERPOLARIZED [1-13C]PYRUVATE<br><br>HYPERPOLARIZED [2-13C]PYRUVATE |

| | | | | | HYPERPOLARIZED [1-13C]PYRUVATE + [13C,15N]-UREA |
|---|---|---|---|---|---|
| **(0018,1040)** | LO | Contrast/Bolus Route | | Administration route of contrast agent | IV |
| **(0018,1041)** | DS | Contrast/Bolus Volume | | Injected volume (mL) | |
| **(0018,1042)** | TM | Contrast/Bolus Start Time | | Define start of injection relative to data acquisition (Not DICOM compliant) Start Time in HHMMSS.FFFFFF (DICOM compliant) | |
| **(0018,1047)** | DS | Contrast/Bolus Flow Duration | | Total injection duration [s] | |
| **(0018,1048)** | CS | Contrast/Bolus Ingredient | IODINE BARIUM GADOLINIUM CARBON DIOXIDE | $^{13}$C enriched compounds | [1-^13^C]PYRUVATE [2-^13^C]PYRUVATE [^13^C,^15^N]UREA [1-^13^C]ALPHA-KETOGLUTARATE |
| **(0018,1049)** | DS | Contrast/Bolus Ingredient Concentration | | Concentration of each $^{13}$C compound [mg/mL] | |
| **(0400,0550)** | SQ | Modified Attributes Sequence | | Store prior versions of an attributes that were removed or modified | |
| **(0400,0562)** | DT | Attribute Modification DateTime | | Tracked Date and Time when attributes were modified | |
| **(0400,0563)** | LO | Modifying System | | Describes what modified the attributes | SIVIC-HP_agent_DICOM_tool.py |

Table 3: DICOM attributes proposed for addition to HP $^{13}$C MRI study data. This captures the majority but not all the desired parameters listed in Table 1. Note that the "Contrast/Bolus Ingredient" attributes listed can include more than one entry to support multiple simultaneously polarized HP $^{13}$C agents[17] such as pyruvate/urea[3,14]. The proposed $^{13}$C naming conventions are in a style that is consistent with the style used for PET in DICOM.

In DICOM, each attribute or data element is identified by a unique tag (TAG), which is comprised of a group number and an element number. The value representation (VR) is a code that indicates the data type used to encode the value(s) of the attribute. Additionally, the value multiplicity (VM) indicates how many values can be present in the attribute. These three components, work together to provide a comprehensive understanding of the attribute and its role within the DICOM file, allowing for accurate and efficient manipulation and interpretation of the medical imaging data.

We reviewed the DICOM standards definition and found the listed attributes in Table 3 can capture much of the unique HP agent metadata. This includes ingredients, injection timing, and dosing characteristics. These can all be added in a DICOM compliant fashion.

One important feature captured in this proposal is the inclusion of multiple ingredients, which can capture dual-agent ("co-polarized") or multi-agent HP studies[3,14,17]. In these studies, multiple $^{13}$C-enriched compounds are simultaneously hyperpolarized and simultaneously injected. The "Contrast/Bolus Agent Sequence", which can capture use of multiple agents that affect the data in a DICOM object, was not used since when multiple HP compounds are desired, they are combined into a single agent with multiple ingredients. The use of multiple agents is not performed to our knowledge, as it would require rapid dissolution of multiple separate samples following hyperpolarized to inject both before the signal decays.

We also proposed standardized ingredient names that are consistent with the style used in PET. In these, the enriched isotope(s) are denoted with "^" around the nucleus, e.g. "^13^C". We also use a standard chemical naming convention to identify the site of enrichment within the molecule.

We used the PyDicom python toolbox to access the data in DICOM files (Figure 2). PyDicom is a general-purpose DICOM framework whose purpose is to reads and write DICOM attributes. It can also add, delete, and modify any attributes to DICOM objects.

```
path = '/Users/ernestodiaz/Desktop/DICOM/AUC_lactate/Image_001.dcm'    ← reading
ds = pydicom.dcmread(path)                                              ← adding
ds.add_new([0x0010,0x0010],'PN',"Ernesto")                              ← writing
ds.save_as(path)
print(ds)
(0010, 0010) Patient's Name                    PN: 'Ernesto'            ← results
```

Figure 2: This is an example of the PyDicom functions used. It reads a file, then adds an attribute and saves the file.

The basic workflow of the python script created using the packages PyDicom and OS is as follows: User would run the script in the terminal with a parameter of a directory path for DICOMS that need to be modified. The prompts would pop up one by one and the user can write its own inputs or press enter for default values as shown in Figure 3. Once the user has entered the prompts, it will loop through the folder and add all the attributes to each DICOM file.

We have also aimed to capture the modifications made to the DICOM file when the tool is used. For this, we used the Modified Attributes Sequence to describe what attributes we modified. For Modifying System, it shows the name of this tool we used to change the attributes. We used DateTime within the Modified Attributes Sequence to keep track of the date when we modified the attributes, and this shows other researchers when we modified them. This maybe a non-standard use of these attributes.

This tool has been demonstrated on DICOM objects from a metabolite-specific EPI sequence[18] on a GE 3T scanner, reconstructed with a custom implementation based on the GE Orchestra package, DICOM from Bruker 2D CSI scan generated by SIVIC, and a DICOM from a GE echo-planar spectroscopic imaging (EPSI) sequence generated by SIVIC.

The script is available within SIVIC at
https://github.com/SIVICLab/sivic/tree/master/applications/dicom_tools.

Figure 4: The prompts questions that gives option for default values or user input. As well, showing the attributes that has been modified.

```
----Welcome----

Enter A Contrast Agent (Press Enter For Default: HYPERPOLARIZED [1-13C]PYRUVATE):
The Current Administration Route of Contrast Agent is Oral & IV. Would you like to overwrite it? Type Y or N:
Enter The Volume of The Contrast Agent Injected In mL (Press Enter For Default: 0.35):
Enter The Total Injection Duration In Seconds (Press Enter For Default: 12):
Enter The Delay Time From Start Of Injection To Start Of Acquisition In Seconds (Press Enter For Default: 0):
Enter A Contrast Ingredient (Press Enter For Default: [1-^13^C]PYRUVATE):
Enter The Molarity Of The Contrast Ingredient in mM (Press Enter For Default: 80):
Enter The Molar Mass Of The Contrast Ingredient in g/mol (Press Enter For Default: 88.06):
Would You Like To Add Another Ingredient? Type Y or N:
IM-0001-0241.dcm
These are the attributes that you modified
[(0400, 0562) Attribute Modification DateTime    DT: '2024031623'
 (0400, 0563) Modifying System                   LO: 'SIVIC-HP_Agent_DICOM_Tool_Test.py']
```

# HP $^{13}$C MRI Data Storage Best Practices

We have also created a proposed set of Data Storage Best Practices. These were based upon a survey of research groups at our institution, performing both human and preclinical HP $^{13}$C imaging studies at our institution. The proposed set of Data Storage Best Practices aim to ensure that the stored data completely describes the study as is necessary for subsequent processing. This should also be done so that future researchers can easily retrieve and analyze the data without the need to search multiple sources of information.

Storage of essential study data should be done in a centralized location with high-quality back up, using a single folder per study, including:

- HP spectroscopy/image data (DICOM recommended)
- HP raw data
- $^{1}$H image data (DICOM recommended)
- Copy of any processing scripts
- Metadata – any other information about the study. This can either be directly noted in the data folder, or in a separate database/spreadsheet. At a minimum this should include
    - Image acquisition data (ideally in DICOM files):
    - Flip angle information, especially for variable flip angle or metabolite-specific flip schedules
    - Injection timing characteristics – length of injection, timing of imaging relative to injection(See Table 1)
    - Dose characteristics – volume injected, concentration of agent, subject weight, polarization.(See Table 1)

- Other Study data - directly recorded in the study folder, or in a separate database/spreadsheet.
- Study "Key" needed in all data locations to link to other metadata sources.
    - Option 1: Unique ID for each study
    - Option 2: Unique ID for each subject + Unique ID for each study

## Discussion and Unmet Needs

Using the DICOM standard has significant advantages for HP $^{13}$C MRI data. Given this technology is being translated into clinical studies, DICOM will allow for seamless integration into existing image viewing and analysis platforms, ultimately supporting use in clinical workflows. The use of standardized DICOM objects would also be valuable in the multi-site clinical trials setting, which are likely to begin within the next several years.

We also believe that the DICOM format will be valuable for the preclinical studies given the flexibility of the format. A pre-clinical DICOM standard has recently been proposed[19], and a recent assessment preclinical imaging metadata need has also advocated for integration into DICOM[20]. While adoption is less consistent in preclinical imaging systems, we have demonstrated a pipeline for our Bruker preclinical MRI scanner.

While the Contrast/Bolus module in DICOM supported most HP agent parameters we wanted to save, there were several parameters without an existing attribute. These include the polarization measurements, T1 relaxation rate, and dissolution timing. These could be stored in private data elements. This could also justify the need for a HP Agent MRI Module in DICOM. This could largely follow the examples from PET, which also must capture similar information: polarization is analogous to specific activity, relaxation rates are analogous to half-life, and dissolution timing is analogous to dose creation timing. The PET Isotope Module (C.8.9.2) could serve as a good model.

Several other current features of the DICOM standard may also be relevant for HP MRI. The Enhanced MR Information Object Definitions (IODs) are attractive because they allow for more MRI specific data sources such as raw k-space and spectroscopy data and can be used for "raw" images (e.g. dynamic metabolite images) as well as derived parameter maps (e.g. kinetic rate maps). The Enhanced Contrast/Bolus Module may also be relevant to store additional HP agent parameters.

Hyperpolarization is also used in HP $^{129}$Xe MRI, a modality that unique images pulmonary ventilation, gas exchange, and terminal airway morphology[22] and received FDA approval in 2021. There are many shared aspects of both HP MRI methods, including the hyperpolarization of an agent, administration of this agent (inhaled for HP $^{129}$Xe), relatively rapid imaging of the agent, and measuring agent kinetics. With this in mind, we believe it would be beneficial to coordinate efforts across the HP modalities when developing DICOM integration methods and specifications of metadata.

HP $^{13}$C MRS and MRSI acquisition methods are more challenging to use in a standardized fashion compared to imaging-based methods (e.g. metabolite-specific imaging). This is because the adoption of DICOM MRS by vendors is rather limited, leading to the community development of tools such as SIVIC[8]. If MRS and MRSI methods continue to show value, it would be important for researchers to work closely with vendor partners and ideally have the DICOM MRS standard integrated into the scanners.

Ultimately, we believe it is important that both a standardized and essential set of DICOM attributes are recorded for HP $^{13}$C MRI studies. This should be standardized so as not to introduce confusion, for example due to different naming conventions or use of different DICOM attributes. This should also include an essential set of attributes to allow for robust and reproducible analyses of HP $^{13}$C MRI data. The final definitions are beyond the scope of this paper, but we think would be best addressed by a consensus of HP $^{13}$C MRI researchers in future discussions as the community recently formed the HP 13C Consensus Group[2]. Following consensus, any changes to the DICOM standard should then be proposed by drafting a Correction Proposal (CP) and work with the DICOM MR Working Group (WG 16) to submit it to the base standards WG (WG 6).

## Conclusion

We have proposed a set of attributes to store HP $^{13}$C MRI metadata, particularly regarding the HP agent, and implemented via PyDicom a program to add these a set of these attributes in a DICOM-compliant fashion based on user input. This was shown to work in line with a variety of DICOM creation pipelines that are used for HP $^{13}$C data. In addition, we have proposed a broader set of data storage best practices. Providing further data standardization will advance this technology by allowing sites to seamlessly share and compare data, including in the context of multi-site trials, and by using DICOM HP $^{13}$C MRI can be easily integrated into other clinical and research workflows.